\documentclass[notitlepage]{article}
\usepackage{natbib}
\usepackage{graphicx}

\begin{document}

\title{Theory and simulations of air shower radio emission}

\author{T. Huege\\
  \small IKP, Karlsruhe Institute of Technology (KIT)
}

\maketitle

\begin{abstract}
A precise understanding of the radio emission from extensive air showers is
of fundamental importance for the design of cosmic ray radio detectors as
well as the analysis and interpretation of their data. In recent years,
tremendous progress has been made in the understanding of the emission
physics both in macroscopic and microscopic frameworks. A consistent picture
has emerged: the emission stems mainly from time-varying transverse currents
and a time-varying charge excess; in addition, Cherenkov-like compression of
the emission due to the refractive index gradient in the atmosphere can lead
to time-compression of the emitted pulses and thus high-frequency contributions in the signal. In this article, I discuss the
evolution of the modelling in recent years, present the emission physics as it is understood today, and conclude
with a description and comparison of the models currently being actively developed.
\end{abstract}


\section{Introduction}

Radio emission from cosmic ray air showers has been measured for the first time in the 1960s \citep{JelleyFruinPorter1965}. This started a flurry of activities both on the experimental and theoretical side; for a review of this early history, please see \citep{Allan1971}. However, difficulties in the interpretation of the measurements and limitations of the analogue detection technique led to a complete cease of activities in the 1970s.

The field was revived in the early 2000s by the LOPES \citep{FalckeNature2005} and CODALEMA \citep{ArdouinBelletoileCharrier2005} experiments. It was clear from their start that a sound understanding of the radio emission physics was imperative for the interpretation of measurements and the optimization of the experimental designs. Therefore, in parallel to the revival of experimental activities, also theorists started new efforts for the modelling of air shower radio emission.

Here, I give an overview of these ``modern'' efforts. I shortly review how the models have evolved and describe the approaches currently employed. Afterwards, I give a summary of the current understanding of the radio emission physics and conclude by presenting some results gathered by a task force aimed at comparing the existing models in detail.


\section{The evolution of modern radio emission modelling}

A driver for the modern efforts in radio detection of cosmic ray air showers has been the paper by Falcke \& Gorham \citep{FalckeGorham2003}. It predicted that air shower radio emission should be measurable with digital radio detectors, and presented a first estimate of the field strengths to be expected on the basis of the ``geosynchrotron'' approach. 
\subsection{Obsolete approaches}
As a direct follow-up to this paper, the geosynchrotron concept was followed up with semi-analytic calculations in the frequency domain \citep{HuegeFalcke2003a} and with Monte Carlo calculations in the time domain \citep{SuprunGorhamRosner2003}. In parallel, full-fledged Monte Carlo simulation codes were developed, in particular REAS1 \citep{HuegeFalcke2005a,HuegeFalcke2005b} and ReAires \citep{DuVernoisIcrc2005}. REAS1 made the transition to REAS2 \citep{HuegeUlrichEngel2007a} by replacing the parameterized air shower model with one based on detailed CORSIKA-derived histograms. Other semi-analytical calculations followed \citep{MeyerVernetLecacheuxArdouin2008,ChauvinRiviereMontanet2010}. For a review of the evolution of these models in the time from 2003 to 2009, please see \citep{HuegeArena2008}.

However, it turned out that all of these models were employing an incomplete modelling of the radio emission physics. They did not take the radiation associated with the varying number of charges in air showers into account correctly, and therefore gave inconsistent results. A particular signature of this inconsistency was that the models always predicted unipolar time-domain pulses, whereas other approaches predicted bipolar pulses. For a detailed discussion of the cause for and solution of these discrepancies, I kindly refer the reader to \citep{HuegeLudwigScholtenARENA2010}. Readers should be aware that the models listed in this section should be considered obsolete and should thus no longer be actively used. Still, many of the predictions that were made by these models, e.g. on coherence effects \citep{HuegeFalcke2003a}, air shower geometry \citep{HuegeFalcke2005b}, and energy and depth of shower maximum sensitivity \citep{HuegeUlrichEngel2008} were qualitatively correct.
\subsection{Current approaches} \label{sec:currentmodels}
With the understanding of the inconsistencies in the models, the field has made tremendous progress. Nowadays, many different approaches exist, and they all agree in their qualitative description of the radio emission features. The approaches can be grouped in the two categories of \emph{microscopic} and \emph{macroscopic} models. The microscopic models follow individual particles (electrons and positrons) in an air shower and calculate the associated radio emission. Macroscopic models rather describe the radio emission physics on the basis of macroscopic quantities such as currents and net charge. Both have their virtues and limitations.
\subsubsection{Microscopic approaches}
A number of microscopic models exist today. The formalisms used to calculate the electromagnetic radiation differ, but they have in common that no assumptions are made on the actual ``radio emission mechanism''. The electromagnetic radiation is calculated on the basis of classical electrodynamics applied in a specific formalism. The radiation that is emitted is thus directly governed by the movement, and in particular acceleration, of the individual particles. (In this sense, it might be more appropriate to call these approaches ``simulations'' rather than ``models''.) The following is a list of existing and actively maintained microscopic models with a short description for each of them. (Other microscopic models exist but are no longer actively maintained and thus not listed here.)
\begin{itemize}
\item {REAS3.1 \citep{LudwigHuege2010} uses the endpoint formalism \citep{JamesFalckeHuege2012} to calculate the radio emission on the basis of histogrammed CORSIKA showers (the same as in REAS2) in the time-domain. }
\item {SELFAS2 \citep{MarinRevenu2012} calculates the radio emission on the basis of a formalism separating the electric field into components for the static contribution, charge variation and current variation in the time-domain. Its air shower model is based on the universality of histogrammed CORSIKA simulations \citep{LafebreEngelFalcke2009}, so no air shower simulation is needed for a radio simulation.}
\item {CoREAS \citep{HuegeARENA2012a} employs the endpoint formalism as in REAS3.1, but calculates the radio emission directly in CORSIKA on a per-particle level, without any intermediate histogramming step. It also works in the time-domain.}
\item {ZHAireS \citep{AlvarezMunizCarvalhoZas2012} employs the ZHS algorithm \citep{ZasHalzenStanev1992} on a per-particle level in AIRES in either the time- or the frequency-domain.}
\end{itemize}
All of these models include a realistic treatment of the atmospheric refractive index. The two ``full Monte Carlo'' simulations CoREAS and ZHAireS describe the underlying air shower with the highest degree of complexity. 
\subsubsection{Macroscopic approaches}
In contrast to the microscopic models, which calculate the emission from individual air shower particles and in principle make no assumptions on the emission ``mechanism'', the macroscopic models calculate the emission as emanating from the ``bulk features'' of the particle distributions, such as currents, net charge or dipole moments. Their advantage is that they are faster and give useful insights in the emission physics. On the other hand, in some cases free parameters have to be set, and it is sometimes difficult to separate the different contributions \citep{JamesFalckeHuege2012}.
\begin{itemize}
\item {MGMR \citep{ScholtenWernerRusydi2008} calculates the radio emission on the basis of time-varying transverse currents, time varying net charge and a time-varying dipole moment in the time-domain. The air shower model is based on parameterizations, some important parameters such as the drift velocity of the particles and the shower-disk thickness need to be chosen.}
\item {EVA \citep{WernerDeVriesScholten2012} employs the same time-domain emission calculation as MGMR, but parameterizes the air shower particle distributions on the basis of individual CONEX simulations. It thus has no free parameters to set. The parameterizations are currently one-dimensional, but could be generalized to multiple dimensions.}
\item {The semi-analytical model developed by Dave Seckel \citep{SeckelARENA2012} employs a similar approach as MGMR, but works in the frequency domain.}
\end{itemize}
Of these models, EVA also includes a realistic treatment of the atmospheric refractive index.


\section{The emission physics}

Having shortly reviewed the models available and actively maintained today, it is instructive to discuss our current understanding of the radio emission physics.
\subsection{Identified contributions}
\begin{figure}[h!t]
\centering
\includegraphics[width=0.7\textwidth]{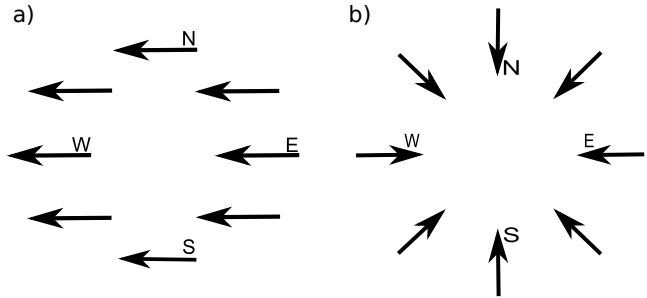}
\caption{Alignment of the electric field vectors as a function of observer position (shower core at the center) for a vertical air shower. Left: transverse current emission, right: charge-excess emission. Both contributions are linearly polarized.} \label{fig:polpatterns}
\end{figure}

\begin{figure}[h!t]
\centering
\includegraphics[width=0.7\textwidth]{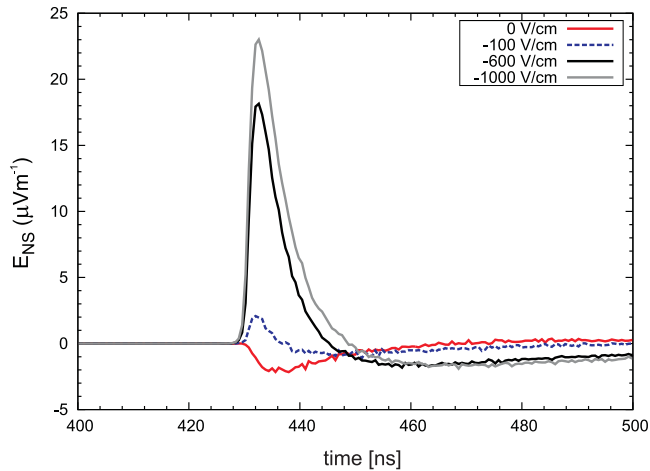}
\caption{North-south component of the electric field for an observer 250~m north of the core of a 30$^{\circ}$ inclined 10$^{16}$ eV air shower simulated at the LOPES site with CoREAS 1.0 in the presence of various vertical atmospheric electric fields.} \label{fig:efieldeffects}
\end{figure}

While disentangling different ``mechanisms'' for the radio emission from extensive air showers (or any complex radiating system) is difficult and sometimes even misleading \citep{JamesFalckeHuege2012}, it is still often instructive to try. For air shower radio emission, the following contributions could be identified:
\begin{itemize}
 \item{The dominant contribution to the radio emission is of geomagnetic origin. The emission is linearly polarized with the electric field vector aligned in the direction of the Lorentz force ($\vec{v} \times \vec{B}$), independent of the observer location as shown in Figure \ref{fig:polpatterns}a). The radiation can be understood as emanating from a time-varying transverse current. The current is a consequence of the equilibrium between acceleration of electrons and positrons in the geomagnetic field and deceleration by interactions with atmospheric molecules. The time-variation arises from the growth and decline of the number of charges during the air shower evolution. This is the mechanism that was already proposed in the 1960s by Kahn \& Lerche \citep{KahnLerche1966}.}
\item{A second contribution to the radio signal arises from the time-variation of the net charge present in an air shower. This is the same mechanism responsible for radio emission of showers in dense media. In air, however, the contribution is only sub-dominant with respect to the geomagnetic emission. The emission is linearly polarized with the electric field vectors oriented radially with respect to the shower axis. Consequently, the polarization angle varies with observer location as presented in Figure \ref{fig:polpatterns}b). The net charge in an air shower arises mainly because electrons are knocked out of atmospheric molecules and are swept along with the shower disk. It is time-varying also because of the growth and decline of the number of charges in the air shower. This is the emission mechanism proposed by Askaryan \citep{Askaryan1962a,Askaryan1965}.}
\item{To which extent the originally proposed ``geosynchrotron'' emission, i.e.\ the radiation associated directly with the acceleration of electrons and particles in the geomagnetic field, contributes to the radio signal is not yet fully resolved; it is conceptually difficult to disentangle completely from the geomagnetic effects leading to the transverse currents. It might play a role at very high frequencies, where recent simulations have predicted polarization characteristics similar to what has been predicted in the earlier ``geosynchrotron'' models \citep{HuegeARENA2012a}.}
\item{It has been known for a long time that atmospheric electric fields can influence the radio emission from air showers, and the effects have also been studied theoretically \citep{BuitinkHuegeFalcke2010}. In recent modelling efforts, it could be shown that, depending on the relative orientation of electric field and shower axis, the transverse current and charge excess contributions can be amplified or dampened by atmospheric electric fields, see Figure \ref{fig:efieldeffects}.}
\item{An effect first noticed by Dave Seckel in REAS3 simulations is a secondary pulse arising at later times for suitable air shower geometries. This arises from the ground impact of the particles, which are stopped instantaneously when reaching the observation level in the REAS3 simulation. A more realistic treatment should take into account the electric properties of the soil and model the ``transition radiation'' arising from this ground impact. This could well be relevant at kHZ frequencies.}
\item{Some macroscopic models include other contributions such as a time-varying dipole or the ions which are left behind in the wake of the air shower. This might be seen as an illustration of the complexity of ``summing up mechanisms''.}
\end{itemize}
In summary, the main mechanisms for the radio emission are the time-varying transverse currents and the time-varying net charge present in the air shower. The superposition of these differently polarized contributions directly leads to prominent asymmetries in the radio emission footprint (in particular an east-west asymmetry for vertical air showers), as is obvious from Figure \ref{fig:polpatterns}.
\subsection{Refractive index effects}
Another effect is important for the radio emission from extensive air showers. The refractive index of the atmosphere is not unity but has a value of $\approx 1.000292$ at sea level and decreases with height as the atmospheric density decreases. The fact that the index is larger than unity means that for suitable geometries radiation emitted at different times and locations can reach a given observer at the same time. In such situations, the pulses are thus compressed in time, and consequently the frequency spectra of the emitted radiation can extend up to very high frequencies well in the GHz regime \citep{AlvarezMunizCarvalhoZas2012,DeVriesBergScholten2011,WernerDeVriesScholten2012}. Many such high-frequency results have recently been presented \citep{HuegeARENA2012a,CarvalhoARENA2012,ScholtenARENA2012}, and they all agree qualitatively in their prediction of a ``Cherenkov ring'' appearing at high frequencies. The diameter of this ring is related to the geometrical distance between source and observer, and thus carries information on the air shower evolution. While it is adequate to label these refractive index effects ``Cherenkov effects'', this should not be mixed up with what is usually referred to as classical ``Cherenkov radiation'' \citep{JamesFalckeHuege2012}.


\section{Where we stand today}

The modelling of radio emission from extensive air showers has come a long way, and with the resolution of the discrepancies present in the first ``modern'' approaches, a consistent picture has emerged. The goal of the modelers should be to understand the radio emission on a level on par with the systematic uncertainties of existing and future experiments, which means on a $\approx 10\%$ level.
\subsection{Model-model comparisons}
\begin{figure}[h!t]
  \includegraphics[width=\textwidth]{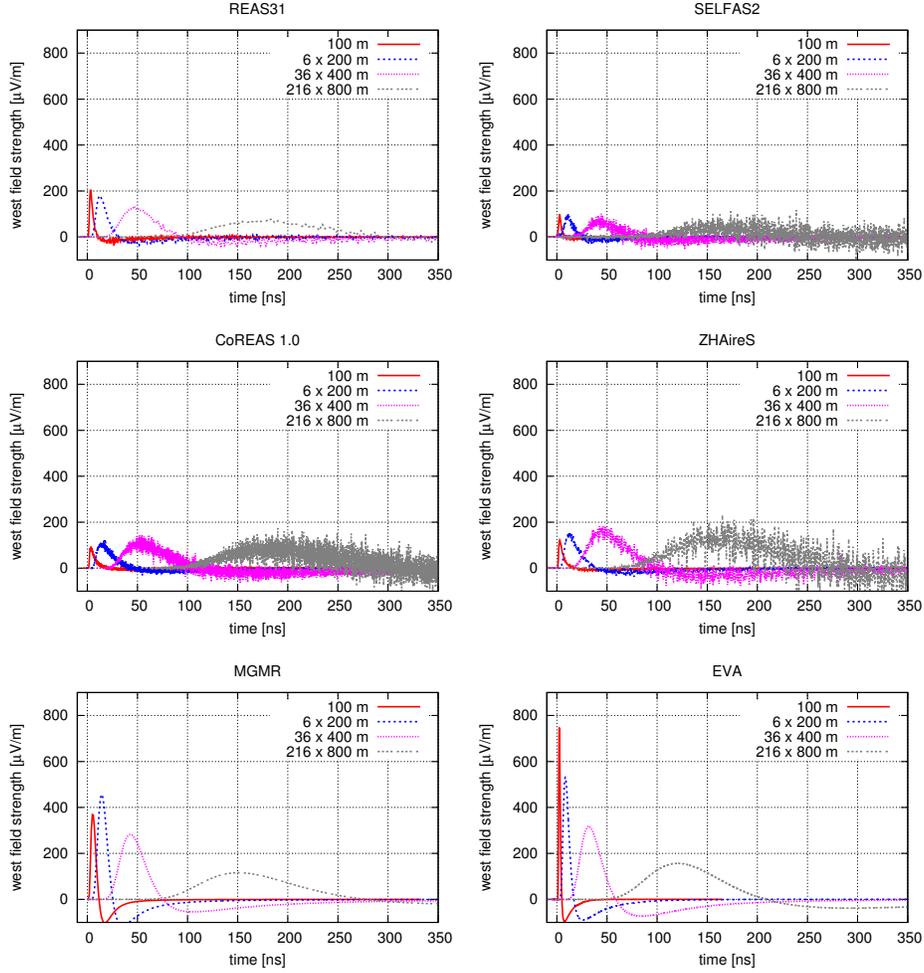}
  \caption{West-component of the electric field pulses for a vertical $10^{17}$~eV proton-induced air shower at the site of the Pierre Auger Observatory as simulated with the indicated models. The refractive index of the atmosphere was set to unity.} \label{fig:comppulsesn1}
\end{figure}

\begin{figure}[h!t]
  \includegraphics[width=\textwidth]{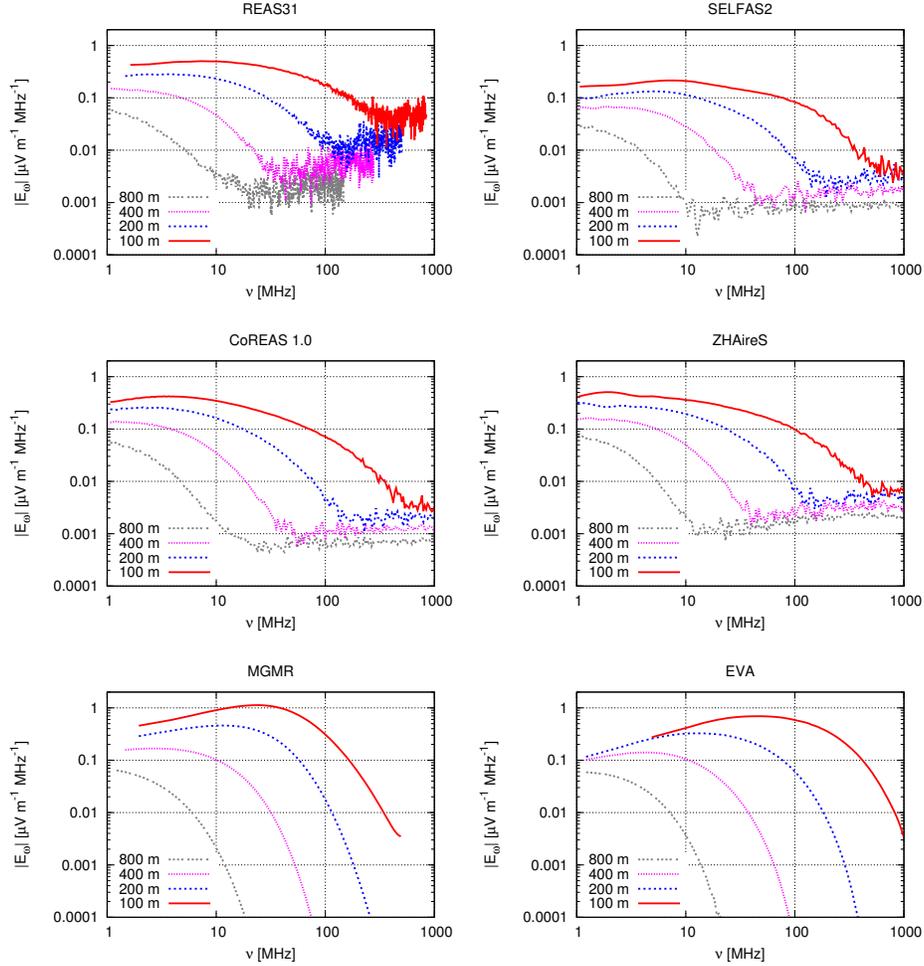}
  \caption{Absolute amplitudes of the frequency spectra for a vertical $10^{17}$~eV proton-induced air shower at the site of the Pierre Auger Observatory as simulated with the indicated models.  The refractive index of the atmosphere was set to unity.} \label{fig:compspectran1}
\end{figure}

\begin{figure}[h!t]
  \includegraphics[width=\textwidth]{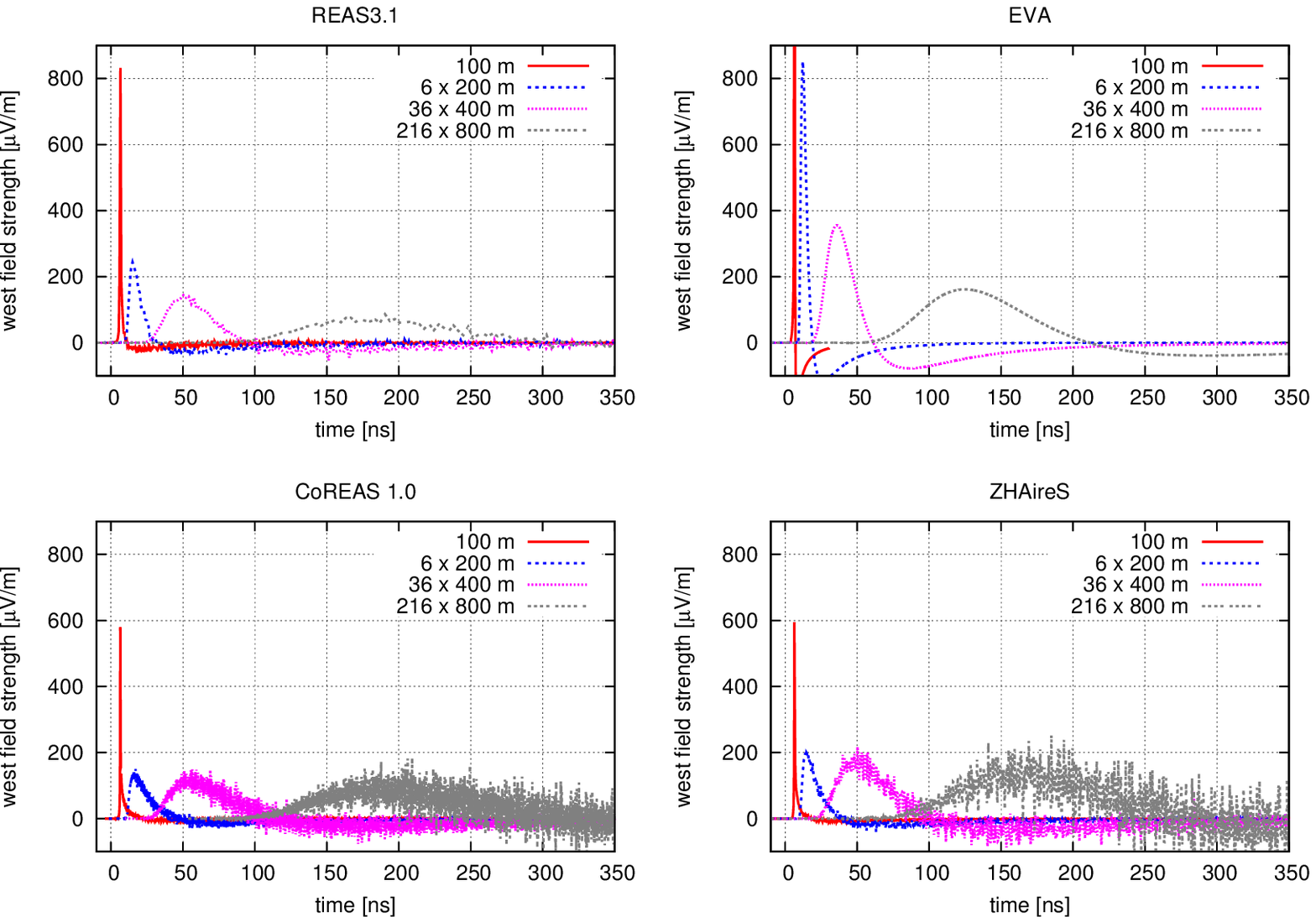}
  \caption{West-component of the electric field pulses for a vertical $10^{17}$~eV proton-induced air shower at the site of the Pierre Auger Observatory as simulated with the indicated models. The refractive index of the atmosphere was modelled according to atmospheric density.} \label{fig:comppulsesnr}
\end{figure}

One way of judging how close the models are to reaching this goal is to compare them systematically. A task force involving authors of all models discussed in section \ref{sec:currentmodels} has started to undertake this effort, initiated at a workshop at Ohio State University in early 2012. In preparation for the ARENA conference, these comparisons were updated once more, and I have the honor of presenting a small excerpt of these on behalf of the task force in Figures \ref{fig:comppulsesn1}, \ref{fig:compspectran1} and \ref{fig:comppulsesnr}.

In these figures, it is striking that the results from CoREAS and ZHAireS are very similar. These two codes model the interaction and emission physics with the highest complexity, i.e.\ on a microscopic per-particle level, and are independent in their technical implementations of both the air shower simulation (CORSIKA vs.\ AIRES) and the radio emission calculation (endpoint formalism vs.\ ZHS algorithm). Their correspondence is thus a very promising indication that the models are indeed converging. REAS and SELFAS show slight deviations from the full Monte Carlos, which might be related to the fact that they are based on information from histogrammed particle distributions rather than individual particles. Another striking observation is that the macroscopic MGMR and EVA models predict significantly higher amplitudes, especially for observers close to the shower axis. These differences to the microscopic models are not yet understood.


%
\subsection{Model-data comparisons}
While comparisons between different models are important and studying the reasons for deviations will be imperative to further improving the models, it is comparisons with data which matter in the end. Unfortunately, currently no data set is publicly available for the modelers to compare their models against. It would be of great use if experiments could publish a data set with all necessary details (including not only air shower parameters such as energy and angles but also the exact core positions and relative antenna locations) so that modelers can use them to benchmark their predictions. Until then, comparisons carried out by the experimental collaborations \citep{LudwigARENA2012} will be the only way to judge the models.
\subsection{Open questions}
One issue that bothered modelers over the past year was the question whether the established calculations worked correctly very near the Cherenkov angle of individual particle tracks. A number of contributions to these proceedings have addressed this issue and the preliminary conclusion is that for the geometries relevant for air showers, the treatments seem to be valid \citep{JamesARENA2012,BelovARENA2012}. Other aspects that should be studied in more detail than they have before include the transition radiation arising from the particle ground impact, the influence of realistically modelled atmospheric electric fields and possible reflection or scattering of radio waves on the particle plasma in the shower disk itself.

\section{Conclusions}

The modelling of radio emission from cosmic ray air showers has made tremendous progress in the past few years. A consistent picture has emerged, and several microscopic and macroscopic models are available and agree in their qualitative predictions. In the next few years, the models have to be benchmarked against each other and against measured data, with the goal to reach and prove a modelling accuracy within the systematic uncertainties of experiments.


\section*{Acknowledgments}
I would like to thank very much all colleagues who participated in the coordinated comparison of models of which I had the honor of presenting a small excerpt at the ARENA conference, in particular: J. Alvarez-Mu\~niz, W. Carvalho Jr., M. Ludwig, V. Marin, B. Revenu, D. Seckel and K.D. de Vries. Also, I would like to thank M. Gelb for his work on modelling electric field effects.



\bibliographystyle{aipproc}   


\IfFileExists{\jobname.bbl}{}
 {\typeout{}
  \typeout{******************************************}
  \typeout{** Please run "bibtex \jobname" to optain}
  \typeout{** the bibliography and then re-run LaTeX}
  \typeout{** twice to fix the references!}
  \typeout{******************************************}
  \typeout{}
 }

\end{document}